# Transport Current in MgB$_2$ based Superconducting Strand at 4.2 K and Self-Field

M.D. Sumption[1], X. Peng[1], E. Lee[1], M. Tomsic[2], and E.W. Collings[1]

[1] LASM, MSE, The Ohio State University, Columbus, OH, 43210
[2] Hyper Tech Research, Tipp City, OH

**Abstract**

Transport current values of 7.5 x 10$^4$ A/cm$^2$ at 4.2 K and self-field are reported for MbB$_2$-based tapes. MgB$_2$ strands were formed by directly filling commercially available MgB$_2$ powder into Nb-lined, monel tubes and then wire drawing. The wires were then rolled into tapes 2.56 x 0.32 mm$^2$, with a total superconducting cross section of 0.2319 mm$^2$. Transport measurements were performed using a standard four-point technique at $T$ = 4.2 K (in liquid helium) and at self field. Three samples were prepared, with heat treatments of 900°C for 1, 2, and 3 h under 1/3 at Ar. Measured values of transport current were 4.7, 7.5, and 1.1 x 10$^4$ A/cm$^2$, respectively, at 4.2 K and self field. M-H loops taken on the sample HT for 1 h showed magnetic $J_c$s of 4.2 x 10$^4$ A/cm$^2$ at 4.2 K and 1 T, indicating that the material had reasonably well connected grains.

**Introduction**

The recent announcement of the discovery of $MgB_2$ [1] has triggered a great deal of interest in the research community. The origin of its transition has been described in terms of an electron-phonon coupling [2] with its associated isotope effect [3]. Alternatively, it has been described in terms of a hole-carrier-based pairing mechanism [4]. The crystal structure of $MgB_2$ is the AlB2 structure, with honeycomb layers of boron atoms alternating with hexagonal layers of Mg atoms. $MgB_2$ decomposes peritectically, with the phase diagram given in [4]. Even though this makes formation of single crystals difficult, polycrystalline $MgB_2$ can be easily formed. In fact, $MgB_2$ powder is commercially available, and has been successfully sintered into a reasonably well connected pellets by heat treatment (HT) under pressure at 1000°C [5]. It is also possible to start with sufficiently small elemental powders and sinter to form high quality $MgB_2$ by reaction at similar temperatures [6].

A number of groups have now fabricated the powders in pellet or similar form and have reported $B_{c2}$ and magnetically derived $J_c$ properties [5-7]. Takano et al. [5] starting with -100 Mesh $MgB_2$ powders, sintered them at 775, 1000, and 1250°C under pressure. They find some level of MgO (present in the powder as-received) and $MgB_4$ impurity phases (at higher temperatures). The powders are seen to sinter together during the 1000°C HT, leading to 20 K, 1 T magnetic $J_c$s of approximately $10^5$ A/cm$^2$ for the powder and 4 x $10^4$ A/cm$^2$ in the sintered bulk. On the other hand Larbalestier et al. have started with elemental Mg and B powders. They pressed them into pellets, placed them on Ta foil on $Al_2O_3$ boats and fired them under Ar for 1 h at 600°C, 800°C, and 900°C.

The powders were then lightly ground, pressed into pellets and HT under pressure at temperatures between 650°C and 800°C for times between 1 and 5.5 h. They obtained $10^4$ A/cm$^2$ at 20 K and 1 T, and also 4 x $10^4$ A/cm$^2$ at 4.2 K 1 T. Recent results from Dou et al., using again elemental powders gives $10^5$ A/cm$^2$ at 4.2 K, 1 T for the $J_c$ of currents flowing across the whole of the slab 3 x 3 x 2 mm [8].

Larbalestier et al. [6] give values of 17.5 T for $B_{c2}(0)$, similar to the estimates from [5]. The level of intrinsic anisotropy has not been determined. The grain to grain link properties seem good in view of results from [5-7] and also [8]. In fact the macroscopic $J_c$ was sufficiently high in [8] that flux jumping was seen.

The properties of MgB$_2$ are increasing rather quickly, suggesting that this system will be much easier to work with than "high T$_c$" materials. The next step is to make the material in wire form. Below some initial progress in that direction is shown.

**Experimental**

MgB$_2$ powder (-100 mesh) was filled into a Nb-lined monel tube 6 mm in diameter. The tube was then drawn through conical dies to form a wire 50 mils in diameter which was subsequently rolled. The samples were then encapsulated under Ar and reacted at 900°C for 1 and 2 hours. Transport $J_c$ measurements were made in liquid helium at self field using the standard four-point technique. M-H loops were measured using an EG&G PAR vibrating sample magnetometer with a 1.7 T iron core magnet.

## Results

Transport measurements were made for two samples, listed in Table 1, below. Sample MGB1, heated treated 1 had an $I_c$ of 108 A and $J_c$ of 4.7 x $10^4$ A/cm$^2$, while sample MGB2 had an $I_c$ of 173 A and a $J_c$ of 7.5 x $10^4$. MGB2 went normal via quench, indicating that the intrinsic transport $J_c$ is higher, and the practical $J_c$ will be improved with proper stabilization. M-H loops were performed on MGB1, and the result is shown in Fig. 1. Here the loop has a ferromagnetic component due to the monel which causes the unusual looking asymmetry, and a peak at low fields due to Nb. However, we have extracted the resulting 1 T $\Delta M$ and it is listed in Table 2. Magnetic $J_c$ was extracted using the expression $\Delta M = (0.2/3\pi)J_c d$, where $d$ is the width perpendicular to the applied field. For the present measurements the field was applied perpendicular to the sample. The resulting magnetic $J_c$ of 4.2 x $10^4$ A/cm$^2$ at 4.2 K and self field is close (somewhat less) than the transport $J_c$ for this sample. The slightly lower magnetic $J_c$ is due to a slight drop in $J_c$ between self field and 1 T.

## Summary

Transport current values of 7.5 x $10^4$ A/cm$^2$ at 4.2 K and self-field were shown for MbB$_2$-based tapes. MgB$_2$ strands were formed by directly filling commercially available MgB$_2$ powder into Nb-lined, monel tubes and then wire drawing. The wires were then rolled into tapes 2.56 x 0.32 mm$^2$, with a total superconducting cross section of 0.2319 mm$^2$. Transport measurements were performed using a standard four-point technique at $T$ = 4.2 K (in liquid helium) and at self field. Three samples were prepared, with heat treatments at 900°C for 1, 2, and 3 h under 1/3 at Ar. Measured values of transport current were 4.7, 7.5, and 1.1 x $10^4$ A/cm$^2$, respectively, at 4.2 K and self field. M-H loops taken on the sample HT at 900°C/1 h showed magnetic $J_c$s of 4.2 x $10^4$ A/cm$^2$ at 4.2 K and 1 T, indicating that the material had reasonably well connected grains.

**Acknowledgments**

We wish to thank S.X. Dou and J. Horvat from the ISEM in Wollongong, Australia for their helpful comments and discussions.

Table 1. Transport $J_c$ values for $MgB_2$ samples at 4.2 K and self field

| Sample Name | Heat Treatment | Superconductor Area, mm$^2$ | $I_c$ (4.2 K, self field), A | $J_c$ (4.2 K, self field), A/cm$^2$ |
|---|---|---|---|---|
| MGB1 | 900°C/1h | 0.23 | 108 | 4.7 x 10$^4$ |
| MGB2 | 900°C/2h | 0.23 | 173 | 7.5 x 10$^4$ |
| MGB3 | 900°C/3h | 0.23 | 25 | 1.1 x 10$^4$ |

Table 2. Magnetic $J_c$ values for $MgB_2$ samples at 4.2 K and 1 T

| Name | Superconductor, w x t x L mm$^3$ | $\Delta M$, emu/cm$^3$, 1 T, 4.2 K | Magnetic $J_c$, A/cm$^2$ 4.2 K, 1 T |
|---|---|---|---|
| MGB1 | 2.26 x 0.103 x 10.0 | 400 | 4.2 x 10$^4$ |

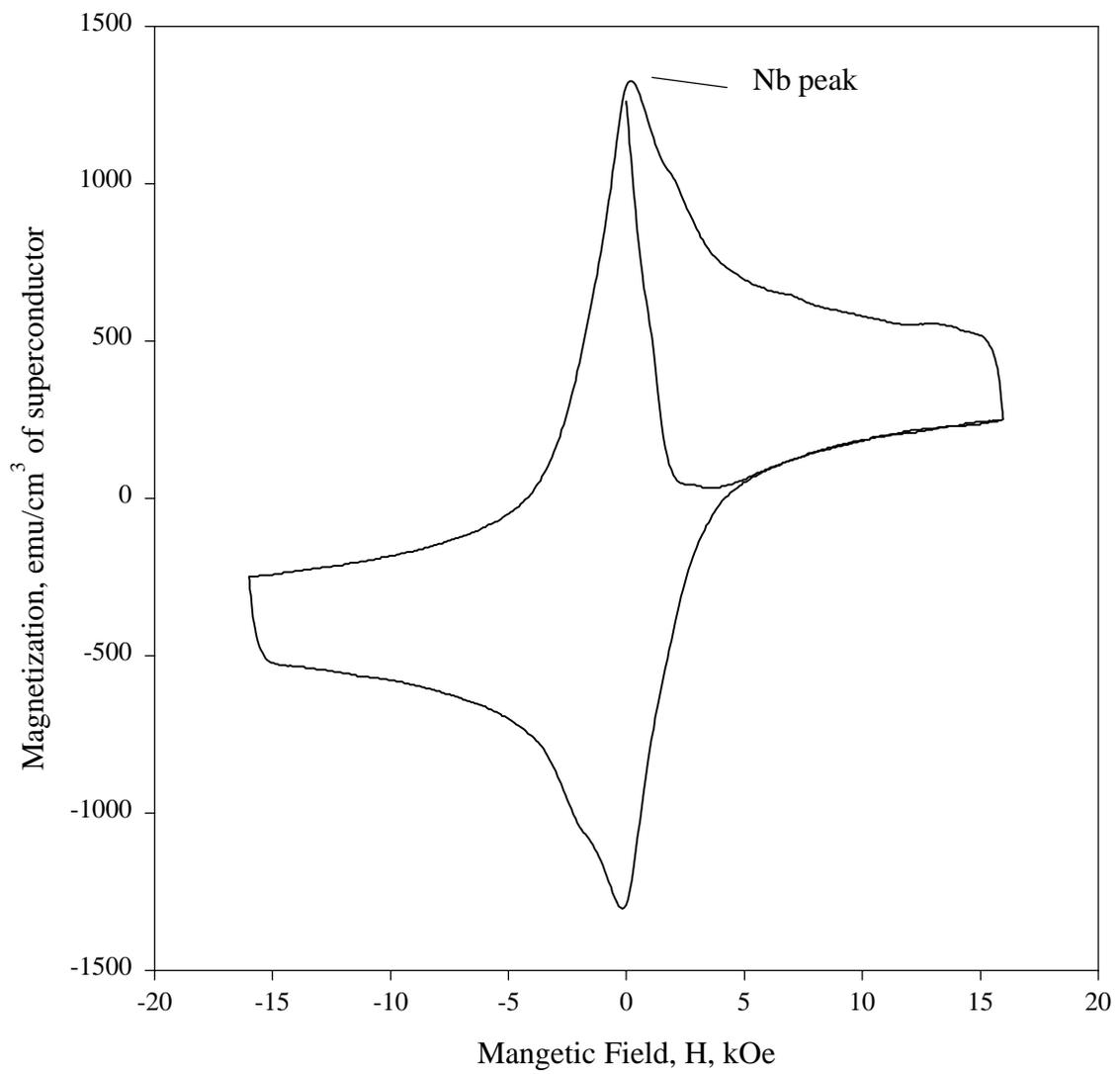

Figure 1. M-H loops for $MgB_2$ sample MGB1 at 4.2 K